\documentclass[12pt,twoside]{article}
\usepackage{amsmath}

\input BoxedEPS
\SetRokickiEPSFSpecial  %% for dvips by Tom Rokicki, for VMS
\HideDisplacementBoxes

\newcommand{\rd}[1]{\mathop{\mathrm{d}#1}}

\newcommand{\fract}[2]{{\textstyle\frac{#1}{#2}}}
\newcommand{\grad}{\vec\nabla}
\newcommand{\vpi}{\vec\pi}
\newcommand{\vp}{\vec p}
\newcommand{\vr}{\vec r}
\newcommand{\vv}{\vec v}
\newcommand{\va}{\vec a}
\newcommand{\vA}{\vec A}
\newcommand{\vB}{\vec B}
\newcommand{\vS}{\vec S}

\newcommand{\CS}{Chern-Simons}

\newcommand{\gi}{gauge-invariant}

\newcommand{\numeq}[2]{\begin{equation}
#2
\label{#1}
\end{equation}}
\newcommand{\refeq}[1]{(\ref{#1})}

\let\vec\boldsymbol
\let\eps\varepsilon
\let\epsilon\varepsilon
\let\phi\varphi

\begin{document}
 
\title{Dirac's Magnetic Monopoles (Again)}
\author{R. Jackiw\\
\small\it Center for Theoretical Physics\\ 
\small\it Massachusetts Institute of Technology\\ 
\small\it Cambridge, MA 02139-4307
}%\\[1ex]
%\small\sl(American Association of  Physics Teachers,   Philadelphia, January 2002)}
\date{}
%\\

\date{\small MIT-CTP-3327}
\maketitle 
\abstract{\noindent
Dirac's quantization of magnetic monopole strength is derived without reference to a (singular,
patched) vector potential.}

\vfill

\centerline{\large Dirac Memorial, Tallahassee, Florida, December 2002}

\thispagestyle{empty}

\newpage

\vspace*{-.5truein}

\section*{}

Dirac's monumental works opened various areas of inquiry in physics and mathematics. His
equation not only became the paradigmatic description for the elementary constituents of
matter, but also was recognized by mathematicians as encoding in its eigenvalues far-reaching
information about geometry and  topology of manifolds. His delta function stimulated the
development of an entire field of mathematics -- the theory of distributions, or generalized
functions. This is how Laurent Schwartz, the creator of that field, put it:
\begin{quote}
I heard of the Dirac function for the first time in my second year at the E[cole] N[ormale]
S[up\'erieure]\ldots. those formulas were so crazy from the mathematical point of view that
there was simply no question of accepting them. \cite{r1}
\end{quote}

My own research, like that of all other physicists, is completely dependent on these
magnificent explorations by Dirac. But there also are other paths that he blazed, which I have
followed. He formulated the quantization of field theory on unconventional surfaces,
corresponding to classical initial value problems posed on these surfaces. This suggested my
 construction (with Cornwall) of light-cone current algebra~\cite{r2}, and (with Fubini and
Hanson) of radial quantization~\cite{r3}, which now is a tool in string theory. Dirac
showed how to quantize dynamical systems that evolve in time while obeying constraints.
Reformulating his approach, in order to simplify it, led Faddeev and me to propose a
Darboux-based solution to the same problem~\cite{r4}. Dirac posited a time-dependent
variational principle, which leads to the time-dependent Schr\"odinger equation. Even though
he didn't seem to publicize it -- it appears only in an appendix to the Russian translation of his
textbook -- Kerman and I used it to define variationally the quantum effective action, and in
an approximate implementation of the  variational principle to derive the time-dependent
Hartree-Fock equations~\cite{r5}. These days his concept of a filled negative energy sea
appears old-fashioned and awkward; mostly it is replaced by normal ordering prescriptions in
quantum field theory. Nevertheless, reference to this apparently unphysical construct gives
the most physical picture for quantum anomalies and for charge fractionation, as was
demonstrated by Feynman~\cite{r6} and Schrieffer~\cite{r7}, respectively. 

A particularly tantalizing result by Dirac  concerns his monopoles. As is well known, he
showed that within quantum mechanics monopole strength has to be quantized, but the
quantization does not arise from a  quantal eigenvalue problem. Rather quantization is
enforced by the requirement that the phase-exponential of the classical action be gauge
invariant. The Lagrangian and the action for motion in a magnetic field are not manifestly
gauge invariant, since they involve the gauge-variant vector potential, rather than the
gauge-invariant magnetic field. Moreover, because the vector potential for magnetic
monopoles is singular, a gauge transformation shifts the action by a constant, and the phase
exponential of the action remains unchanged only when this constant is a proper multiple of
$2\pi$. This then is the origin of Dirac's famous quantization condition, and it has a precise field
theoretical reprise in the quantization of the \CS\ coefficient in odd-dimensional gauge
theories, as was shown by Deser, Templeton, and me~\cite{r8}.

Dirac's quantization argument has been thoroughly scrutinized, and is certainly acceptable.
But one wonders whether one could reach the same conclusion in a gauge-invariant manner,
relying on \gi\ quantities and dispensing with reference to gauge-variant and singular vector
potentials. 

Here I shall present such an argument, which I constructed some years ago~\cite{r9}.
Although it is not new, it is not widely known. Moreover, it not only regains the Dirac
quantization condition, but also demonstrates that quantal magnetic sources must be
structureless point particles. 

Let us begin by recording the \gi\ Lorentz-Heisenberg equations of motion for operators
$\vr(t)$ specifying the motion of a massive ($m$) charged ($e$) particle in an external
magnetic field $\vB$:
\begin{gather}
\dot{\vr} = \vv  \label{e1}\\
m\dot{\vv} = \frac e{2c} \bigl[
\vv\times\vB - \vB\times\vv \bigr]\ . \label{e2} 
\end{gather}
In the second equation, the noncommuting operators $\vv$ and $\vB(\vr)$ are symmetrized.
Since the magnetic field does no work, the conserved energy $\fract12 m\vv^2$ does not see it.
This energy formula also gives us the Hamiltonian~$H$ that generates the above equations by
commutation,
\begin{gather}
H = \frac{\vpi^2}{2m} \qquad \vpi\equiv m\vv \label{e3}\\
\intertext{provided the following brackets are posited:}
[r^i, r^j] = 0 \label{e4}\\
[r^i, \pi^j] = i\hbar\delta^{ij}\label{e5}\\
[\pi^i, \pi^j] = ie\frac\hbar c \eps^{ijk} B^k (\vr)\ . \label{e6}
\end{gather}
Note that $\pi^i$ is not the (gauge-variant) canonical momentum; rather it is the
(gauge-invariant) kinematical momentum. With  \refeq{e3}--\refeq{e6} eqs.~\refeq{e1} and
\refeq{e2} are reproduced as 
\begin{align}
\dot{\vr} &= \frac i\hbar [H, \vr] = \frac\vpi m \label{e7}\\
\dot{\vpi} &= \frac i\hbar [H, \vpi] = \frac e{2mc} \bigl[
\vpi\times\vB - \vB\times\vpi\bigr]\ . \label{e8}
\end{align}

The equations of motion \refeq{e1}, \refeq{e2} or \refeq{e7}, \refeq{e8} do not appear to
require any constraint on~$\vB$. They make sense whether $\vB$ is source free
$\grad\cdot\vB=0$, or not $\grad\cdot\vB\neq 0$. However, when we look to the Jacobi
identity for the commutators of the $\vpi$'s, we find $\fract12 \eps^{ijk}\bigl[
\pi^i, [\pi^j, \pi^k]\bigr] = \frac{e\hbar^2}c \grad\cdot\vB$. This vanishes, as it should, for
source-free magnetic fields, which then are given by the conventional curl of a vector
potential, $\vB= \grad\times\vA$, and   momenta $\vp$ canonically conjugate to $\vr$ realize
the algebra
\refeq{e5},
\refeq{e6} with the formula
\numeq{e9}{
\vpi = \vp - \frac ec \vA(\vr)\ .
}

\begin{figure}[h!]
\centerline{\BoxedEPSF{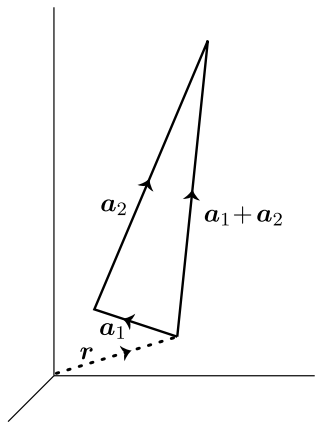 scaled 900}}
\caption{The triangle at $\vr$ through which the flux $\Phi$ is calculated.}\label{diracfig1}
\end{figure}

But how are we to understand the occurrence of magnetic sources with the concomitant
violation of the Jacobi identities? To make progress on this question, recall that commutators in
an algebra state the infinitesimal composition law for the corresponding finite transformations.
In particular \refeq{e5} shows that
\numeq{e10}{
T(\va) \equiv \exp\Bigl( -\frac i\hbar \va\cdot\vpi\Bigr)
}
effects translations by $\va$ on $\vr$:
\numeq{e11}{
T^{-1}(\va)\,\vr\, T(\va) = \vr + \va\ .
}
If the $\vpi$ were commuting momenta, the product of  $T(\va_1)$ with $T(\va_2)$ would
reflect the Abelian composition law of translations and close on $T(\va_1 + \va_2)$. Here,
however, because the $\vpi$'s do not commute [see \refeq{e6}] we find
\numeq{e12}{
T(\va_1) T(\va_2) = \exp \Bigl(-\frac{ie}{\hbar c} \Phi (\vr; \va_1,\va_2) \Bigr)T(\va_1+\va_2)
}
where $\Phi(\vr;  \va_1,\va_2)$ is the magnetic flux through the triangle with vertices $(\vr; 
\vr+\va_1,\linebreak[3]\vr+\va_1+\va_2)$ (in the direction $\va_1\times \va_2$). (See
Fig.~\ref{diracfig1}.)

\begin{figure}[hbt]
\centerline{\BoxedEPSF{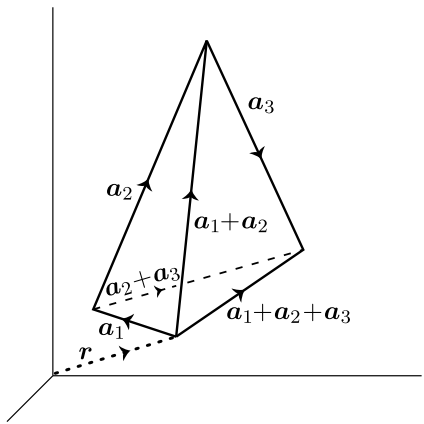 scaled 900}}
\caption{The tetrahedron at $\vr$ through which the flux determining the
nonassociative phase~$\omega$ is calculated.}\label{diracfig2}
\end{figure}

The Jacobi identity is the infinitesimal statement of associativity in the composition law. Its
failure when $\grad\cdot\vB\neq0$ means that finite translations do not associate. Indeed,
from
\refeq{e12} we have, on the one hand,
\begin{align}
\Bigl(T(\va_1) T(\va_2)\Bigr) T(\va_3) &=
 \exp \Bigl(-\frac{ie}{\hbar c} \Phi (\vr; \va_1,\va_2) \Bigr)T(\va_1+\va_2)\, T(\va_3)
\label{e13} \\
&= \exp \Bigl(-\frac{ie}{\hbar c} \Phi (\vr; \va_1,\va_2) \Bigr)
\exp \Bigl(-\frac{ie}{\hbar c} \Phi (\vr; \va_1+\va_2, \va_3) \Bigr) T(\va_1+\va_2 + \va_3)
\nonumber
\end{align}
and on the other
\begin{align}
T(\va_1) \Bigl(T(\va_2)T(\va_3)\Bigr)  &=
T(\va_1) \exp \Bigl(-\frac{ie}{\hbar c} \Phi (\vr; \va_2,\va_3) \Bigr)T(\va_2+\va_3) 
\label{e14} \\
&= \exp \Bigl(-\frac{ie}{\hbar c} \Phi (\vr-\va_1; \va_2,\va_3) \Bigr)
T(\va_1) T(\va_2+\va_3) 
\nonumber\\
&= \exp \Bigl(-\frac{ie}{\hbar c} \Phi (\vr-\va_1; \va_2,\va_3) \Bigr)
\exp \Bigl(-\frac{ie}{\hbar c} \Phi (\vr; \va_2+\va_3) \Bigr)
T(\va_1+\va_2+\va_3) \ .
\nonumber
\end{align}
Putting everything together we find that 
\numeq{e15}{
\Bigl(T(\va_1) T(\va_2)\Bigr) T(\va_3) =
\exp \Bigl(-\frac{ie}{\hbar c} \omega(\vr; \va_1,\va_2,\va_3) \Bigr)
T(\va_1) \Bigl(T(\va_2) T(\va_3)\Bigr) 
}
where $ \omega(\vr; \va_1,\va_2,\va_3)$ is the total magnetic flux emerging out of the
tetrahedron formed from three vectors $\va_i$, with one vertex at~$\vr$:
\numeq{e16}{
\omega=\int \rd{\vS}\cdot\vB = \int \rd r \grad\cdot\vB\ .
}
The last integral is over the interior of the tetrahedron, and of course vanishes for
source-free magnetic fields, but is nonzero in the presence of magnetic sources, leading in
general to the nonassociativity of the translations $T(\va)$. (See Fig.~\ref{diracfig2}.)

But when operators act on a vector or Hilbert space, they necessarily associate. So one cannot
tolerate nonassociativity within the usual quantum mechanical formalism. The only possibility
for nonvanishing $\grad\cdot\vB$ is that its integral is quantized for arbitrary $\va_1$,
$\va_2$, and $\va_3$:
\numeq{e17}{
\int \rd r \grad\cdot\vB = 2\pi \frac{\hbar c}e N\ .
}
Then $\fract e{\hbar c}\omega$ is invisible in the exponent since it is an integer ($N$) multiple
of~$2\pi$.

Equation \refeq{e17} saves associativity, but it places requirements on~$\vB$. First, the
magnetic field must be a (collection of) point source(s), so that \refeq{e17} not lose its
integrality when the $\va_i$ are varied: the source must be either inside or outside the
tetrahedron. Moreover, for each point source of strength~$g$ 
\numeq{e18}{
\vB = g \frac{\vr}{r^3}
}
we have $\grad\cdot\vB = 4\pi g \delta^3(\vr)$ and Dirac's quantization is regained
from~\refeq{e17}:
\numeq{e19}{
\frac {ge}{\hbar c} = \frac N2\ .
}

Finally, with point sources we can also save the Jacobi identity, which now is violated only at
isolated points and these may be excluded from the manifold. 

Thus one has arrived at Dirac's result in a \gi\ manner, without ever introducing a vector
potential with its attendant singularities, patches, etc. It would be interesting to know whether
there is a similarly \gi\ derivation for the quantization of the \CS\ coefficient~\cite{r8}.

Noncommutativity skirts what is acceptable mathematics for quantum theory. Its first
manifestation is avoided by Dirac's quantization. Yet noncommutativity has reappeared in
modern string theory. 
It remains to be seen whether mathematical sense can be made of this. Here again we can
appreciate Laurent Schwartz's sentiment,
\begin{quote}
This at least can be deduced\ldots.  It's a good
thing that theoretical physicists do not wait for mathematical justification before going ahead
with their theories. \cite{r1}
\end{quote}

What about the physics, as opposed to the mathematics, of magnetic monopoles? Let me
conclude with Dirac's own assessment:
\begin{quote}
I am inclined now to believe that monopoles do not exist. So many years have gone by without
any encouragement from the experimental side. \cite{r10}
\end{quote}

\small
\def\Journal#1#2#3#4{{\em #1} {\bf #2}, #3 (#4)}
\def\add#1#2#3{{\bf #1}, #2 (#3)}
\def\Book#1#2#3#4{{\em #1}  (#2, #3 #4)}
\def\Bookeds#1#2#3#4#5{{\em #1}, #2  (#3, #4 #5)}
% \Journal{}{}{}{}
% \Book{}{}{}{}

\end{document}